\documentclass[sigplan,10pt]{acmart}
\AtBeginDocument{%
  \providecommand\BibTeX{{%
    \normalfont B\kern-0.5em{\scshape i\kern-0.25em b}\kern-0.8em\TeX}}}

\setcopyright{acmlicensed}
\acmYear{2024}\copyrightyear{2024}
\acmConference[EuroMLSys '24]{4th Workshop on Machine Learning and Systems}{April 22, 2024}{Athens, Greece}
\acmBooktitle{4th Workshop on Machine Learning and Systems (EuroMLSys '24), April 22, 2024, Athens, Greece}
\acmDOI{10.1145/3642970.3655839}
\acmISBN{979-8-4007-0541-0/24/04}

\usepackage[utf8]{inputenc} 
\usepackage[T1]{fontenc}    
\usepackage{hyperref}       
\usepackage{url}            
\usepackage{booktabs}       
\usepackage{nicefrac}       
\usepackage{xcolor}         
\usepackage{natbib}
\usepackage{multirow}
\usepackage{pifont}
\usepackage{graphicx}
\usepackage{subfigure}
\usepackage{bbm}
\usepackage{booktabs} 
\usepackage{mathrsfs}

\usepackage{setspace}
\usepackage{graphicx}
\usepackage{subfigure}
\usepackage{fancyhdr}
\usepackage{hyperref}
\usepackage{microtype}
\usepackage{xcolor}

\pdfoutput=1
\newcount\comments  
\newcommand{\genComment}[2]{\ifnum\comments=1{\textcolor{#1}{\textsf{\footnotesize #2}}}\fi}

\usepackage{wrapfig}
\usepackage{setspace}
\usepackage{graphicx}
\usepackage{subfigure}
\usepackage{booktabs} 

\usepackage{hyperref}
\usepackage{algorithmic}
\usepackage{algorithm}

\usepackage{graphicx}
\usepackage{subfigure}
\usepackage{booktabs} 
\usepackage{setspace}
\usepackage{graphicx}
\usepackage{subfigure}
\usepackage{booktabs} 
\usepackage{graphicx}
\usepackage{subfigure}
\usepackage{bm}
\usepackage{mathrsfs}

\usepackage{hyperref}

\begin{document}

\title{IA2: Leveraging Instance-Aware Index Advisor with Reinforcement Learning for Diverse Workloads}


\author{Taiyi Wang}
\affiliation{%
  \institution{University of Cambridge}
  \city{Cambridge}
  \country{United Kingdom}}
\email{Taiyi.Wang@cl.cam.ac.uk}

\author{Eiko Yoneki}
\affiliation{%
  \institution{University of Cambridge}
  \city{Cambridge}
  \country{United Kingdom}}
\email{eiko.yoneki@cl.cam.ac.uk}




\begin{abstract}



This study introduces the \textbf{I}nstance-\textbf{A}ware \textbf{I}ndex \textbf{A}dvisor (IA2), a novel deep reinforcement learning (DRL)-based approach for optimizing index selection in databases facing large action spaces of potential candidates. IA2 introduces the Twin Delayed Deep Deterministic Policy Gradient - Temporal Difference State-Wise Action Refinery (TD3-TD-SWAR) model, enabling efficient index selection by understanding workload-index dependencies and employing adaptive action masking. This method includes a comprehensive workload model, enhancing its ability to adapt to unseen workloads and ensuring robust performance across diverse database environments. Evaluation on benchmarks such as TPC-H reveals IA2's suggested indexes' performance in enhancing runtime, securing a 40\% reduction in runtime for complex TPC-H workloads compared to scenarios without indexes, and delivering a 20\% improvement over existing state-of-the-art DRL-based index advisors.

\end{abstract}





\maketitle

\section{Introduction}
For more than five decades, the pursuit of optimal index selection has been a key focus in database research, leading to significant advancements in index selection methodologies \cite{lum1971optimization}. However, despite these developments, current strategies frequently struggle to provide both high-quality solutions and efficient selection processes \cite{kossmann2020magic}.

The Index Selection Problem (ISP), detailed in Section \ref{sec:ISP}, involves choosing the best subset of index candidates, considering multi-attribute indexes, from a specific workload, dataset, and under given constraints, such as storage capacity or a maximum number of indexes. This task, aimed at enhancing workload performance, is recognized as NP-hard, highlighting the complexities, especially when dealing with multi-attribute indexes, in achieving optimal index configurations \cite{lan2020index}.

\begin{figure}[htbp]
    \centering
    \includegraphics[width=1.0\linewidth]{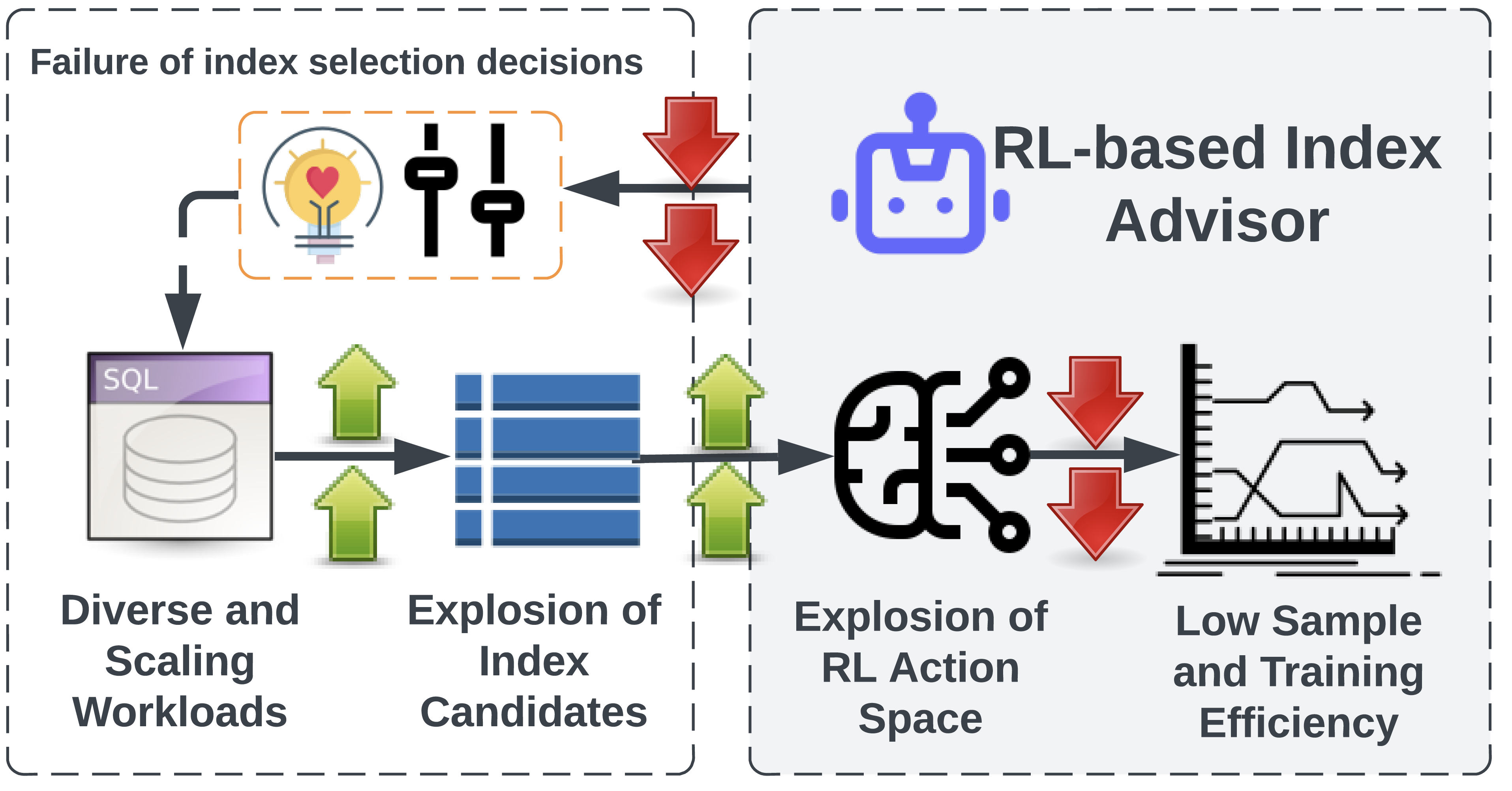}
    \caption{Unique challenges to RL-based Index Advisors due to diverse and complex workloads}
    \vspace{-8pt}
    \label{fig:intro}
\end{figure}

Reinforcement Learning (RL) offers a promising solution for navigating the complex decision spaces involved in index selection ~\cite{kossmann2022swirl,sadri2020drlindex, lan2020index}. Yet, the broad spectrum of index options and the complexity of workload structures complicate the process, leading to prolonged training periods and challenges in achieving optimal configurations. This situation highlights the critical need for advanced solutions adept at efficiently managing the complexities of multi-attribute index selection ~\cite{kossmann2022swirl}. Figure ~\ref{fig:intro} illustrates the difficulties encountered with RL in index selection, stemming from the combinatorial complexity and vast action spaces. Our approach improves DRL agent efficiency via adaptive action selection, significantly refining the learning process. This enables rapid identification of advantageous indexes across varied database schemas and workloads, thereby addressing the intricate challenges of database optimization more effectively.

Our contributions are threefold: (i) modeling index selection as a reinforcement learning problem, characterized by a thorough system designed to support comprehensive workload representation and implement state-wise action pruning methods, distinguishing our approach from existing literature. (ii) employing TD3-TD-SWAR for efficient training and adaptive action space navigation; (iii) outperforming state-of-the-art methods in selecting optimal index configurations for diverse and even unseen workloads. Evaluated on the TPC-H Benchmark, IA2 demonstrates significant training efficiency, runtime improvements, and adaptability, marking a significant advancement in database optimization for diverse workloads.

\section{Related Works}
\label{sec:Related Works}

This section outlines the landscape of existing research on index selection approaches, with a particular focus on traditional index advising methods and reinforcement learning (RL)-based strategies for index advising. Our discussion aims to contextualize the innovations our work introduces in the field.

\subsection{Traditional Index Selection Approaches}




Traditional index selection methods, despite their evolution over decades, often struggle with the intricate interdependencies between indexes and the dynamic nature of workloads and tend to struggle to deal with the explosion of index candidates' choices. Early two-stage greedy-based approaches by Chaudhuri et al.~\cite{chaudhuri1997efficient} and Valentin et al.~\cite{valentin2000db2} made significant strides but failed to consider critical interactions among different indexes, key for optimizing database performance. Similarly, while ILP formulations~\cite{papadomanolakis2007integer} and \textit{Cophy}~\cite{dash2011cophy} brought mathematical precision to modeling the Index Selection Problem (ISP) as a Binary Integer Problem, they too overlooked the complex interplay between indexes and the multifaceted access patterns in contemporary databases.

Among traditional index selection algorithms, \textit{Extend} ~\cite{schlosser2019efficient} represents a significant contribution, characterized by its novel recursive strategy that complements its additive approach to building index configurations. This method stands out by not preemptively excluding index candidates and effectively managing index interactions, addressing the limitations of existing approaches for large database instances. Unlike reductive methods, which often lead to prohibitive runtimes or suboptimal solutions by limiting the set of index candidates early in the process, \textit{Extend} prioritizes both efficiency and solution quality. This approach reflects a broader trend in index advising, seeking to balance the demands of complex analytical workloads with the practical necessities of runtime and scalability.

\subsection{RL-based Index Selection Approaches}

Recent advancements have seen the application of Reinforcement Learning (RL) to the index selection problem, offering novel approaches that promise to overcome some of the limitations of traditional methods.

\textit{DRLinda}, introduced by Sadri et al.~\cite{sadri2020drlindex}, targets cluster databases and, while innovative in its focus on such environments, does not support multi-attribute indexes and lacks a public implementation for validation against state-of-the-art methods.

Lan et al.~\cite{lan2020index} propose an RL-based solution capable of identifying multi-attribute indexes. Despite this advancement, their approach does not model workload representation, limiting its ability to generalize to new or unseen workloads and potentially constraining solution quality due to preselected index candidates.

\textit{SWIRL}~\cite{kossmann2022swirl} represents a state-of-the-art index selection method that surpasses both traditional and RL-based approaches by incorporating a detailed workload model and action masking rules, effectively supporting multi-attribute indexes and excelling in generalizing to new query types. However, \textit{SWIRL}'s sophistication comes with challenges, such as high training costs and complexity. Its detailed approach requires significant computational resources and expertise, and its dependence on manually defined pruning rules can limit training efficiency and adaptability in highly variable environments, highlighting the need for further research to improve its practicality and training process.

In conclusion, while traditional and early RL-based methods have laid the groundwork for automated index selection, they often fail to address the complexity of workloads and database environments fully. Our work seeks to fill this gap, offering a comprehensive solution that balances the need for efficiency, adaptability, and high-quality index configurations.

\section{Index Selection Problem}
\label{sec:ISP}

The \textbf{Index Selection Problem (ISP)} is formalized as the task of identifying an optimal index set, \(I^*\), from a set of candidate indexes, \(I\), for a database, \(D\), and its workload, \(W\), to minimize the execution cost, \( \text{Cost}(W, I) \), subject to constraints, \(C\), such as a storage budget. Formally, this can be represented as:
$$
I^* = \arg\min_{I \subseteq I} \text{Cost}(W,I) \quad \text{s.t.} \quad C(I) \leq C_{\text{max}}
$$
where \(C(I)\) denotes the cost associated with the index configuration \(I\), including considerations such as storage, and \(C_{\text{max}}\) represents the maximum allowable cost under the constraints.

\section{Methodology}

In this section, we outline our novel approach to the Index Selection Problem (ISP), significantly advancing the application of Deep Reinforcement Learning (DRL). Our methodology not only frames the ISP within a DRL context but also introduces a groundbreaking RL model, the Twin Delayed Deep Deterministic Policy Gradients-Temporal Difference-State-Wise Action Refinery (TD3-TD-SWAR). Drawing on the innovative work of ~\cite{sun2022toward} and ~\cite{yoon2018invase}, TD3-TD-SWAR is specifically designed to efficiently manage the ISP's complex solution space. A key novelty of our approach is the adaptive action masking mechanism, which accelerates training and sharpens decision-making by filtering out less beneficial actions based on current conditions, achieving optimal index selection with minimal decision steps.

\subsection{Formulation of the DRL Problem}

Deep Reinforcement Learning (DRL) offers a method for sequential decision-making to optimize database index configurations, addressing the Index Selection Problem (ISP) ~\cite{sutton2018reinforcement,lan2020index}. The goal in ISP is for an agent to find an optimal index set \(I^*\) from candidates to reduce workload execution costs. Given budget constraints, the agent strategically sequences new index additions, navigating the dynamic environment to find configurations that improve database performance over time, showcasing ISP as a fundamental DRL challenge that underscores strategic decision-making under constraints.

Key DRL components for ISP are:
\textbf{Agent}: The system designed to learn optimal index configurations to enhance database performance.
\textbf{Environment}: The database system environment for which the optimization of indexes is performed.
\textbf{States $s$} represent the database and workload status, including index configurations, query plans, and metadata, modeled as a multi-dimensional vector. This ensures a comprehensive view for decision-making and adaptability to new workloads. The representations will be decided our well-designed workload model (see Section ~\ref{sec:sys_pre} for more).
\textbf{Actions $a$}: The act of selecting an index to be added to the configuration, with the aim of improving database workload performance.
\textbf{Reward} \(r_t\): Defined for an action at timestep \(t\), it quantifies the performance improvement due to the new index, adjusted by the cost and normalized by the ratio of the additional storage used, as:
    \begin{equation}
    r_t(I^*_t) = \frac{C(I^*_{t-1}) - C(I^*_t)}{C(\emptyset)} \bigg/ \frac{M(I^*_t)-M(I^*_{t-1})}{M(I^*_{t-1})},
    \end{equation}
    where \(C(I^*_t)\) is the workload cost under configuration \(I^*_t\), \(C(\emptyset)\) is the cost without any indexes, and \(M(I^*_t)\) and \(M(I^*_{t-1})\) represent the storage used by the index configuration \(I^*_t\) and \(I^*_{t-1}\), respectively.
\textbf{Policy} \(\pi\): A strategy that maps the current database state to an action, guiding the agent to select indexes that optimally enhance performance.

While basic Deep Reinforcement Learning (DRL) techniques are applicable, the challenge of large action space, i.e., too many index candidates, necessitates an advanced solution. Our Twin Delayed Deep Deterministic Policy Gradients-Temporal Difference-State-Wise Action Refinery (TD3-TD-SWAR) model addresses this by refining action space pruning, ensuring efficiency and precision in tackling the complexities of index selection under storage and training limitations.

\subsection{Instance-Aware Deep Reinforcement Learning for Efficient Index Selection}
\label{method_instanceaware}

\begin{figure}[htbp]
    \centering
    \includegraphics[width=1.0\linewidth]{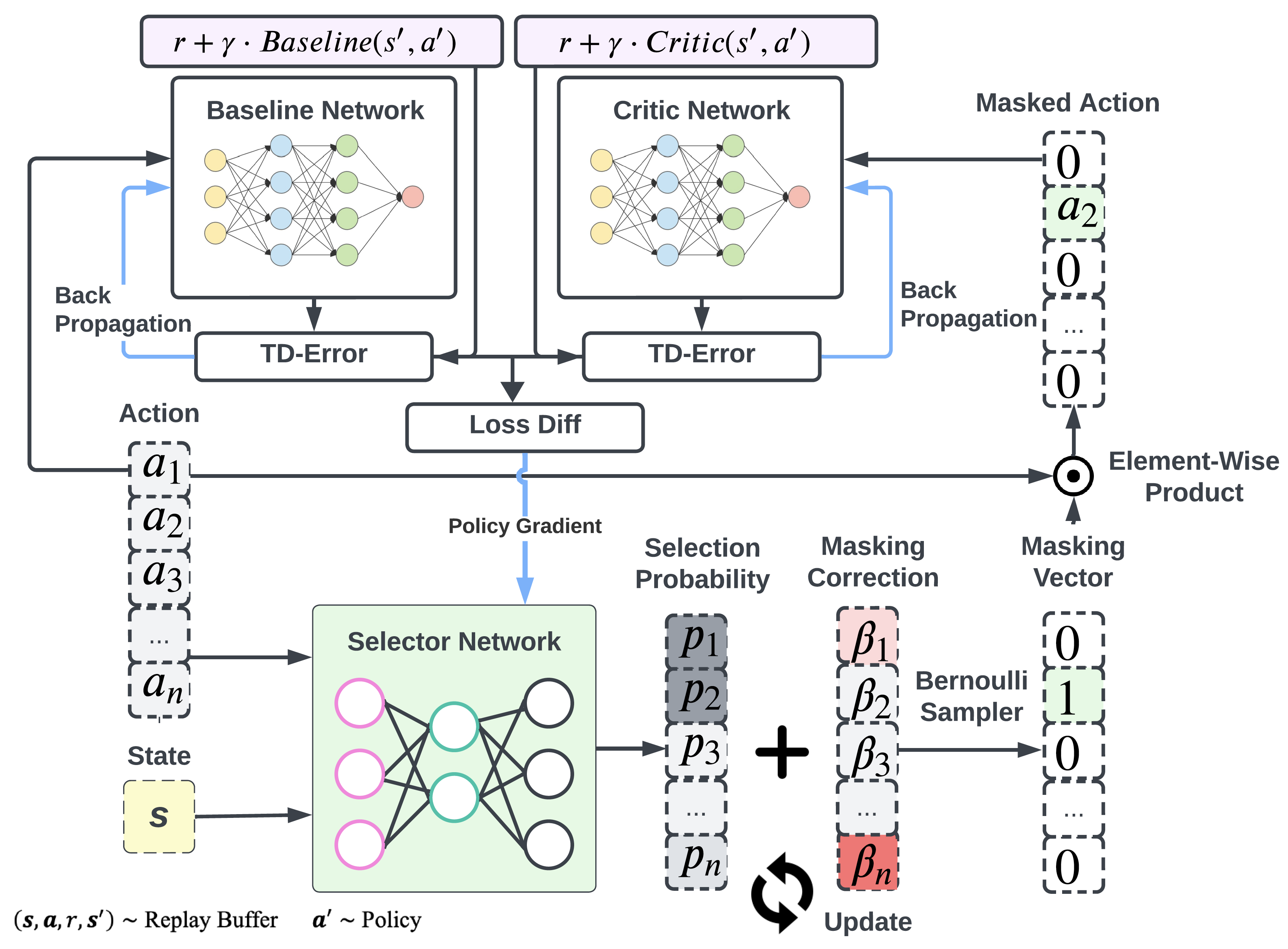}
    \caption{Block diagram of TD3-TD-SWAR in temporal difference learning. States and actions sampled from replay buffer are fed into the selector network that predicts the selection probabilities of different dimensions of actions. A selection mask is then generated according to such a selection probability vector.}
    \label{fig:TD-SWAR}
\end{figure}

Our TD3-TD-SWAR model, developed for the Index Selection Problem (ISP), enhances action space pruning through the incorporation of a selector network (\(G_{\theta}\)). This network selectively masks actions based on their relevance, concentrating on those with substantial impact to improve computational efficiency. The model extends the traditional Actor-Critic reinforcement learning framework ~\cite{konda2000actor}, exemplified by TD3 ~\cite{fujimoto2018addressing}, by adding specific features. One such addition is a blocking diagram that highlights the crucial role of the selector networks, depicted in Figure ~\ref{fig:TD-SWAR}. This approach is grounded in minimizing Temporal Difference (TD) error (\(\mathcal{L}_{TD}\)), directing actors towards the most beneficial actions in accordance with the task's goals:

\begin{equation}
\mathcal{L}_{TD} = \mathbb{E}_{s_i,a_i,r_i, s'_i \sim \mathcal{B}} \left[ \left(r_i + \gamma Q_w(s'_i, \pi(s'_i)) - Q_w(s_i, a_i) \right)^2 \right],
\end{equation}

where the introduction of \(G_{\theta}\) and the blocking diagram in our model signifies our additional design on the existing Actor-Critic RL frameworks like TD3, emphasizing the selector networks' role in optimizing the action selection process.

Training of \(G_{\theta}\) exploits the TD error differences between baseline (unmasked actions) and critic networks (masked actions), pinpointing actions' contributions under storage constraints. This discrepancy informs \(G_{\theta}\)'s refinement, employing policy gradients for targeted action exploration:
\begin{equation}
    \min_{G, Q_w} \mathbb{E}_{s_i,a_i,r_i, s'_i\sim\mathcal{B}} \left[ \left(y'_i - Q_w(s_i, a_i^{(G(a_i|s_i))}) \right)^2 \right]  + \lambda |G(a_i|s_i)|_0,
\end{equation}

This process not only reduces computational demand by focusing on essential actions but also dynamically adjusts \(G_{\theta}\), ensuring action selection is closely aligned with ISP's strategic goals, thereby improving learning efficiency and decision quality. Basic ideas of this proposed RL algorithm are presented in Algorithm ~\ref{AlgorithmCompact}.

\begin{algorithm}[htbp]
\caption{TD3 with TD-SWAR and Enhanced Masking}
\label{AlgorithmCompact}
\begin{algorithmic}\small
    \STATE Initialize networks: critic $C_{\phi_{1,2}}$, baseline $B_{\psi_{1,2}}$, actor $\pi_{\nu}$, selector $G_{\theta}$, and their targets
    \STATE Initialize replay buffer $\mathcal{B}$, mask history $M_{\text{hist}}$, and action existence set $A_{\text{exist}}$
    \FOR{$t = 1$ to $H$}
        \STATE Store transition $(s,a,r,s')$ in $\mathcal{B}$
        \STATE Sample mini-batch from $\mathcal{B}$
        \STATE Generate perturbed action $\tilde{a}\leftarrow \pi_{\nu'}(s') + \text{noise}$
        \STATE Adjust action selection probabilities using $M_{\text{hist}}$ and $A_{\text{exist}}$
        \STATE Mask actions using $G_{\theta'}$ and biases, calculate target values $y_c$, $y_b$
        \STATE Update critic and baseline networks using \textbf{MSE} loss
        \STATE Refine mask history and action set based on current masking
        \STATE Update $G_{\theta}$ with policy gradient; update $\pi_{\nu}$ with deterministic policy gradient
        \STATE Soft update of target networks with rate $\tau$
    \ENDFOR
\end{algorithmic}
\end{algorithm}

\begin{table*}[htbp]
  \centering
  \caption{RL-based index advisors comparison}
  \label{tbl:comparison}
  \begin{tabular}{lcccc}
    \toprule
                                    & DRLinda & Lan et al. & SWIRL & IA2 \\
    \midrule
    Multi-attributes                    & No      & Yes       & Yes   & Yes \\
    Stop criterion                  & \#Idx   & \#Idx,Storage  & Storage    & Storage \\
    Workload representation                  & Yes     & No        & Yes   & Yes \\
    Gen. to new queries             & ++      & -         & +++   & +++ \\
    Training difficulty             & ++      & +         & +++   & + \\
    Action Space                    & Raw     & Rule-based gen.  & Rule-based mask. & Rule-based gen. + Adaptive mask. \\
    \bottomrule
  \end{tabular}

\end{table*}

\vspace{-10pt}

\section{System Framework of IA2}

\begin{figure*}[htbp]
    \centering
    \includegraphics[width=1.0\linewidth]{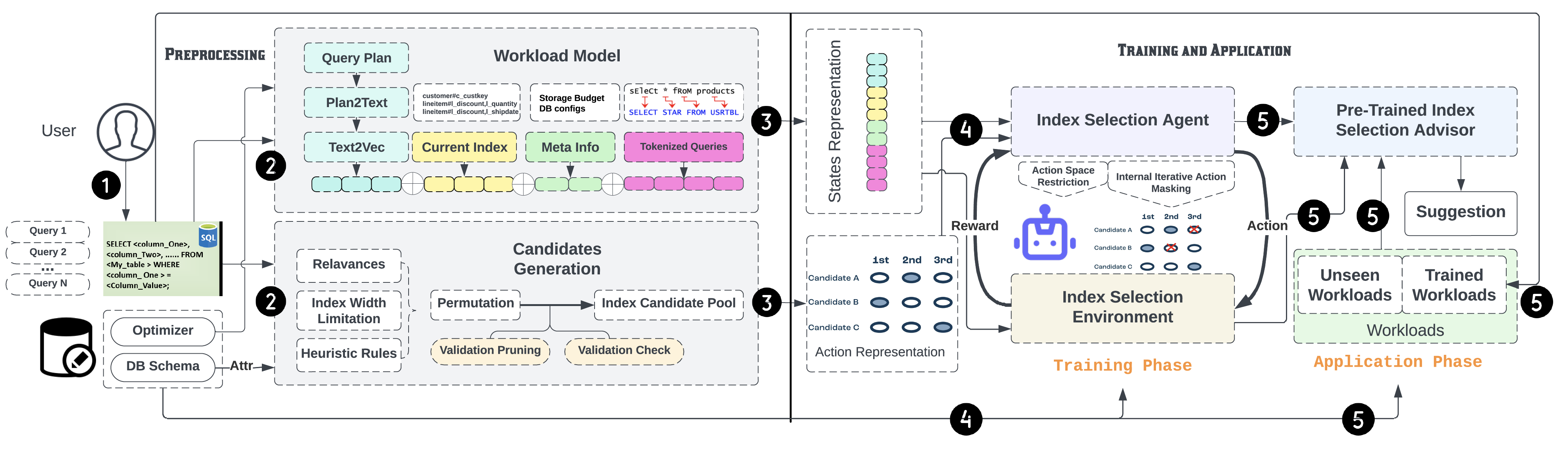}
    \caption{Framework and working flow of IA2}
    \label{fig:IA2-Sys}
\end{figure*}

As shown in Figure ~\ref{fig:IA2-Sys}, IA2 operates through a structured two-phase approach, leveraging deep reinforcement learning to optimize index selection for both trained workloads and unseen scenarios. It depicts IA2's workflow, where the user's input workload is processed to generate states and action pools for downstream RL agents. These agents then make sequential decisions on index additions, adhering to budget constraints, demonstrating IA2's methodical approach to enhancing database performance through intelligent index selection.

\subsection{Preprocessing Phase}
\label{sec:sys_pre}
The preprocessing phase is critical for establishing a solid foundation for IA2's operation. It consists of two components:

\textbf{Workload Model:} Enhanced by the underlying optimizer and what-if cost models, the workload model captures database workload variabilities. It integrates four essential components: Query Plan features, reflecting database reactions; current index configurations; Meta information about database configurations and budget; and embedded tokenized queries. This model is crucial for providing accurate state representations to the downstream DRL training task, significantly boosting IA2's generalization capabilities across diverse workloads.

\textbf{Index Candidates Enumerator:} This component extends beyond exhaustive enumeration, employing validation rules and restrictions to discern the relevance among queries. By leveraging permutations and heuristic rules that cater to generic operators and workload structures, the Enumerator crafts index candidates. This approach, inspired by and integrating advancements from Lan et al. ~\cite{lan2020index} , enriches the selection pool with a broader array of indexing strategies, poised to optimize performance across varying scenarios. The generated index candidates form the raw action space for the downstream RL task, laying a foundational step for IA2's decision-making process in selecting optimal indexes.

\subsection{RL Training and Application Phase}

The RL Training and Application Phase of IA2 transitions from initial preprocessing to actively engaging with defined action spaces and state representations, marked by:

\textbf{TD3-TD-SWAR Algorithm Application:} Leveraging the action space and state representations crafted in the preprocessing phase, IA2 employs the TD3-TD-SWAR algorithm, as outlined in Algorithm \ref{AlgorithmCompact}. Unlike merely operating on preprocessed data, this approach integrates action space restrictions—accounting for existing index candidates and their masking history. Each tuning step recalibrates masking possibilities for subsequent selections, embodying a strategy that adaptively masks actions irrelevant based on the current agent states. 

\textbf{Adaptation to Workloads:} Designed for flexibility, IA2 applies learned strategies to a range of workloads, efficiently adapting to both familiar and unseen environments, demonstrating its capability to handle diverse operational scenarios.





\section{Experiments}
Our experiments are designed to evaluate the IA2 on several critical aspects of database optimization and index selection. Specifically, we aim to (1) analyze the performance of IA2's core algorithm, TD3-TD-SWAR, against other reinforcement learning algorithms, showcasing its unique strengths and contributions; (2) assess the efficiency of the action masking technique in IA2 for action space reduction and learning process acceleration; and (3) measure the end-to-end (E2E) workload runtime improvements achieved with IA2, highlighting its practical impact on database performance.

\subsection{Experimental Setting}

\textbf{Implementation and Environment:} Our prototype is implemented in Python, utilizing PyTorch for model development. Interfaced with PostgreSQL 15.6, it integrates HypoPG for what-if analysis, aiding in query cost estimation. Experiments are conducted on a virtual machine powered by a shared Nvidia Quadro RTX8000 GPU and equipped with 8 CPU cores, within a single-threaded SQL-DB environment.

\begin{figure}[htbp]
    \centering
    \includegraphics[width=1.1\linewidth]{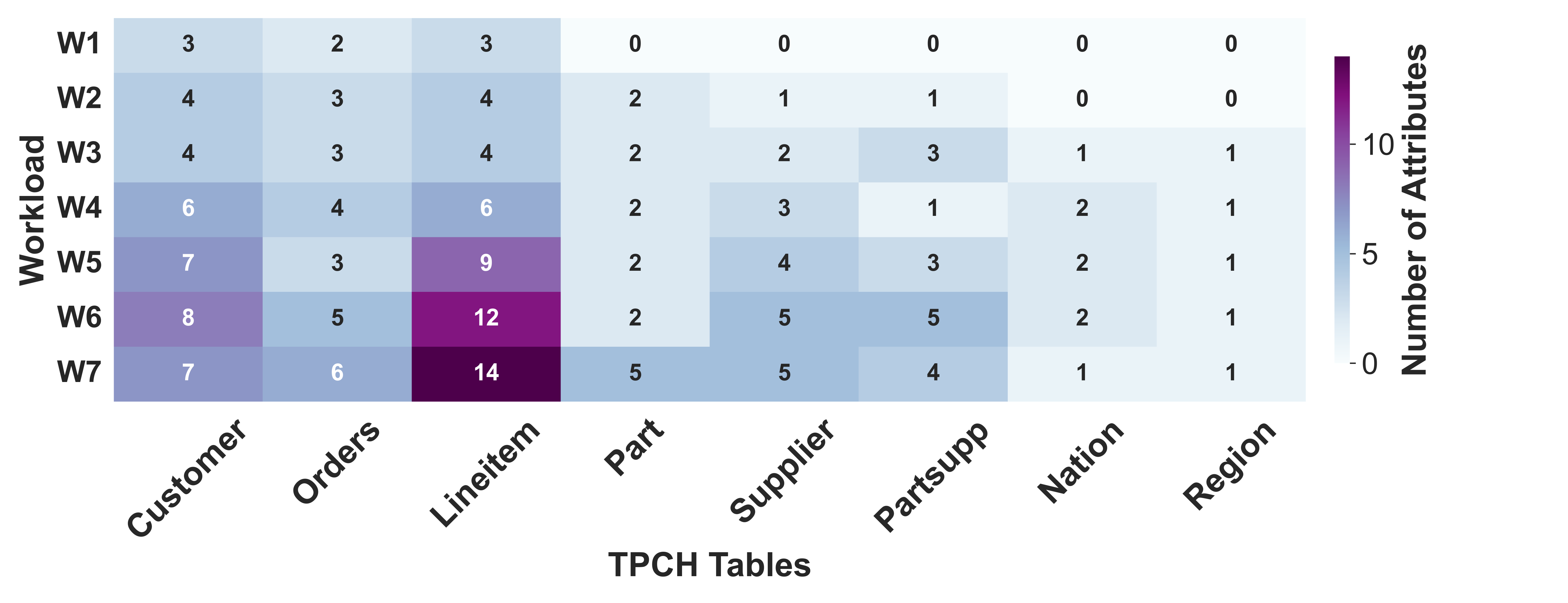}
    \caption{Workloads' Outline, W1-W7 with the increasing complexity and diverse patterns}
    \vspace{-9pt}
    \label{fig:workloads}
\end{figure}

\textbf{Benchmark Workloads:} TPC-H (SF1) forms the basis for seven workloads (W1 - W7), derived from its 22 query templates plus additional queries for a broad evaluation scope. Each workload contains 50 queries, with complexity reflected in the diversity of tables and attributes. W7, uniquely, serves as a test for IA2’s ability to generalize, being unseen during training. W1-W6 are used for standard training and evaluation, while W7 undergoes slight fine-tuning on a subset of the training set for performance assessment on novel queries. Workload outlines are depicted in Figure ~\ref{fig:workloads}.

\textbf{Competitors:} Our evaluation includes comparisons with \textit{SWIRL}, \textit{DRLinda}, \textit{Extend}, and Lan et al., as discussed in Section \ref{sec:Related Works}, to benchmark IA2 against the state-of-the-art in index selection. Comparison of these selected RL-based index advisors is shown in Tabel ~\ref{tbl:comparison}

\textbf{Evaluation Metrics:} The primary metric for assessing IA2 and its competitors is the end-to-end runtime of workloads, using the performance gain ratio for direct optimization comparisons across index advising methods. The evaluation covers trends in Storage Budget (2-8), Workloads, and Training Episodes (50-400), with storage quantified in units where 1 unit equals 128MB.

\subsection{Experimental Results}
\begin{figure}[htbp]
    \centering
    \includegraphics[width=1.0\linewidth]{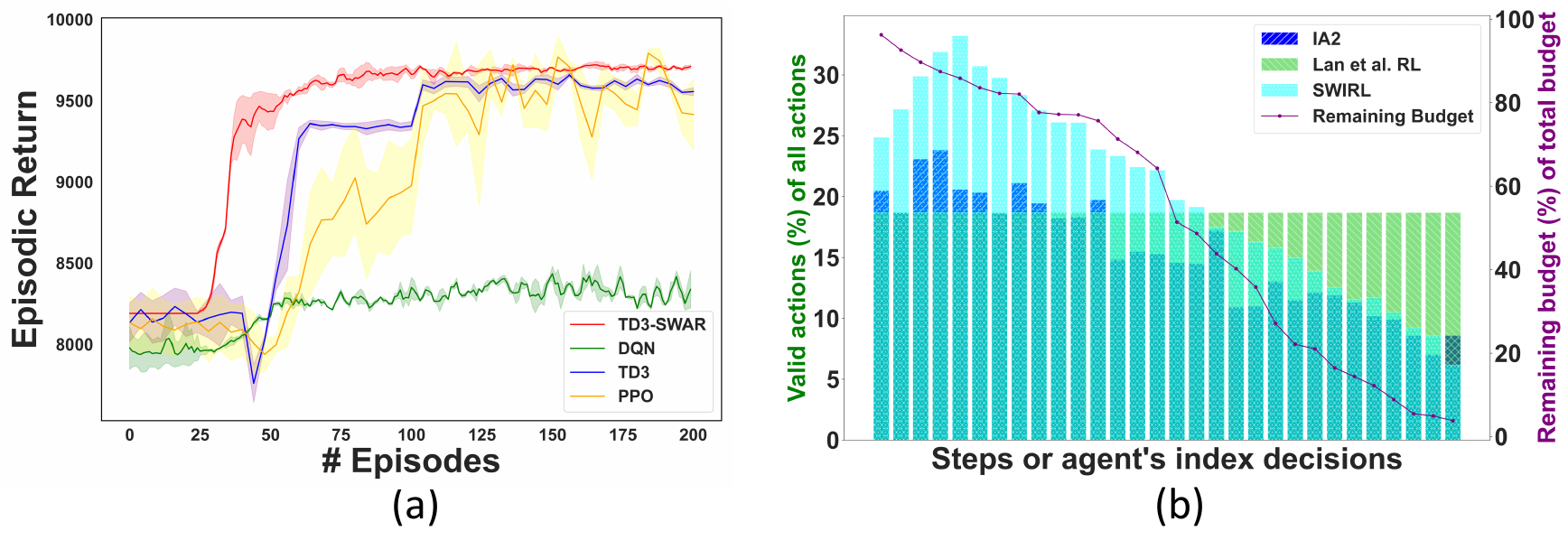}
    \caption{Comparative Analysis of IA2's (a) TrainingEfficiency and (b) Action Pruning Efficiency.}
    \label{fig:combined_figures}
\end{figure}

\paragraph{Training Efficiency of TD3-TD-SWAR:}
In Figure ~\ref{fig:combined_figures} (a), the TD3-TD-SWAR algorithm showcases superior training efficiency against other RL algorithms like DQN, PPO, and vanilla TD3, particularly with the complex workload W6 under an 8-unit storage budget. Remarkably, IA2 completes 100 episodes in just 50 seconds, significantly faster than \textit{SWIRL}, which can take several to tens of minutes under similar conditions. This efficiency is attributed to the effective what-if cost model, ensuring cost-efficient scaling for diverse data and workloads. IA2's capacity for rapid training is further enhanced by its support for pre-trained models, offering adaptability without extensive retraining.

\paragraph{Efficiency of Action Pruning Approaches:}
Figure ~\ref{fig:combined_figures} (b) showcases IA2's action masking efficiency when exhausting the storage budget, comparing it with \textit{SWIRL}'s dynamic masking and Lan et al.’s static heuristic rules using workload W6. With an 8-unit budget and 1701 possible actions, IA2 significantly reduces the action space, especially in the early stages of training. This adaptive strategy emphasizes IA2's ability to navigate and prune the action space more effectively, ensuring a streamlined and focused exploration of indexing strategies.

Our analysis reveals key differences in index selection strategies: \textbf{Lan et al.}\cite{lan2020index} adopts a fixed-rule approach for index combinations, ensuring high training efficiency but potentially overlooking valuable index candidates. \textbf{SWIRL}\cite{kossmann2022swirl} utilizes a combination of exhaustive generation and dynamic masking to explore a wider array of actions, though its effectiveness can vary with workload specifics, impacting training efficiency. As shown in Table ~\ref{tbl:comparison}, \textbf{IA2} merges the benefits of both approaches, employing flexible and automatic selection of meaningful actions for training alongside specific rules for generating candidates, enhancing the efficiency and adaptability of index selection.

\begin{figure}[htbp]
    \centering
    \includegraphics[width=1.04\linewidth]{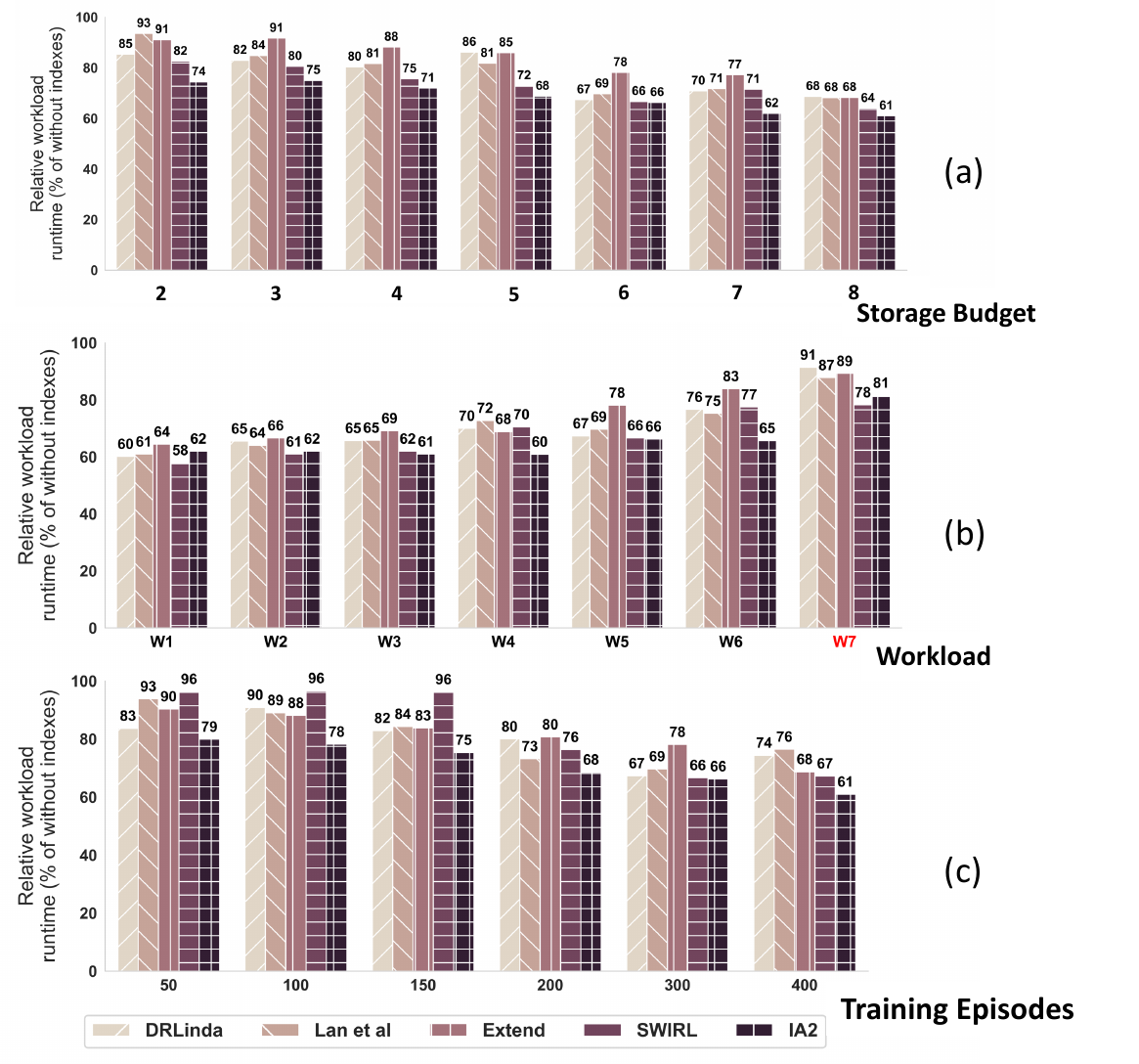}
    \caption{End-to-End runtime performance comparisons across different conditions (\% of runtime W/O Indexes) among Index Advisors: (a) Varying storage budgets with workload W5 over 300 episodes, (b) Differing workloads with a fixed 6-unit storage budget over 300 episodes, and (c) Variations with training episodes for a 6-unit storage budget on workload W5. Though \textbf{Extend} ~\cite{fujimoto2018addressing} is not RL-based, its performance is compared under similar episodic evaluations.}
    \vspace{-6pt}
    \label{fig:runtime}
\end{figure}

\subsection{End-to-End Performance Comparison}

The comprehensive evaluation of index selection methods reveals IA2's distinctive advantages in terms of performance, adaptability, and learning efficiency. This section delves into the comparative analysis across three critical dimensions: storage budget optimization, workload diversity, and training length, highlighting IA2's outperformance. Across various benchmarks, IA2 consistently outperforms other index selection methodologies by an average margin of 15-20\%. This performance differential is not only significant but also indicative of IA2's robust and efficient algorithmic design, which is finely tuned to optimize database query execution times.

\paragraph{Storage Budget Efficiency:}
IA2 demonstrates a remarkable performance gain, with a 61\% improvement over the runtime without indexes (shown in Figure ~\ref{fig:runtime} (a)). This is a significant enhancement compared to other methods, notably \textit{SWIRL}, which peaks at about 64\%. The key differentiation for IA2 lies in its storage-aware RL agent design. By efficiently utilizing the available storage budget, IA2 optimizes index configurations to achieve superior performance gains. Such efficiency is pivotal in scenarios where storage resources are limited, making IA2 a preferred solution for database performance optimization.

\paragraph{Workload Changes:}
Figure ~\ref{fig:runtime} (b) underscores IA2's exceptional adaptability, consistently delivering high performance across complex workloads (W3-W6) and achieving notable improvements in previously unseen scenarios like W7. This demonstrates its robustness and crucial adaptability for dynamic real-world applications.

Conversely, \textit{SWIRL}'s performance improvements on simpler workloads (W1 and W2) and its ability to adapt to the unseen W7 are significantly aided by its intricate workload model, benefiting from its detailed approach to centralized patterns that facilitate action pruning. Nonetheless, these strengths are largely attributed to its elaborate designs and the substantial training resources it consumes. Despite these advantages, IA2 distinguishes itself with superior adaptability and efficiency across a broader spectrum of workloads, affirming its suitability for dynamic environments.

\paragraph{Training Efficiency:}
IA2's training efficiency is a hallmark of its design, achieving optimal performance with fewer training episodes (shown in Figure ~\ref{fig:runtime} (c)). This rapid convergence to peak efficiency is indicative of an efficient learning process, crucial in fast-paced environments where swift adaptation is key. In comparison, \textit{SWIRL}'s performance with limited training underlines the effectiveness of IA2's learning mechanism, which not only conserves time but also computational resources, enhancing cost efficiency.

IA2's training efficiency results in significant operational savings, making it a compelling choice for database optimization, where minimizing training costs without sacrificing performance is crucial. To summarize, IA2 excels in learning efficiency, cost-effectiveness, and adaptability, leveraging storage budgets effectively to boost database performance in environments with limited storage, varied workloads, and a need for swift adaptation, establishing it as a vital tool for database administrators and architects.

\subsection{Key Insights}

Summarizing our extensive experiments, IA2 represents a significant advancement in index selection, outperforming existing methods in several key areas:

\textbf{Rapid Training Efficiency:} IA2 excels with its unparalleled training speed, leveraging a what-if cost model and pre-trained models to facilitate quick adaptability and learning. This efficiency allows IA2 to drastically reduce training time compared to competitors, making it highly suitable for environments where speed is crucial.
    
\textbf{Advanced Workload Modeling:} Unlike static or exhaustive methods, IA2 employs dynamic workload modeling, enabling it to adapt to changing database queries and structures seamlessly. This flexibility ensures optimal index selection across diverse scenarios, including previously unseen workloads.
    
\textbf{Effective Action Space Exploration:} IA2 introduces an innovative approach to pruning and navigating the action space, efficiently identifying meaningful actions early in the training process. This strategy contrasts with the more resource-intensive techniques of SWIRL~\cite{kossmann2022swirl} or the rigid rules of Lan et al.~\cite{lan2020index}, offering a balanced pathway to optimizing index configurations without exhaustive search or oversimplification.

\section{Conclusion and Future Work}
This study introduces the Instance-Aware Index Advisor (IA2), employing the TD3-TD-SWAR model for efficient index selection in databases, showcasing adept handling of complex dependencies and generalization to unseen workloads. Demonstrated through TPC-H benchmarks, IA2 achieves superior efficiency, setting a new standard in index configuration optimization across varied database environments.

Future iterations of this work will aim to expand the discussion on the index choices across IA2 and comparative systems, delving into the nuances of performance differences across various workloads and training epochs. Testing IA2 on a broader set of workloads beyond the TPC-H benchmark and exploring its performance in dynamically changing environments are pivotal steps forward. Such explorations will not only validate IA2’s adaptability and efficiency but also enhance its applicability across diverse database environments. Acknowledging the current evaluation's focus and the limitation in workload diversity, additional evaluations on a more expansive range of real-world workloads and database schemas are planned. Furthermore, exploring compression technologies to enhance IA2's scalability represents a crucial area of development. These future directions aim to broaden IA2's effectiveness and applicability in diverse database scenarios, ensuring its readiness for the dynamic and varied demands of contemporary database systems and paving the way for more resilient, efficient, and intelligent database optimization strategies.

\bibliographystyle{ACM-Reference-Format}
\bibliography{ref}


\end{document}